\documentclass {elsarticle}		

\usepackage{lineno,hyperref}

\journal{Journal of Computational Physics}

\usepackage{amsmath}
\usepackage{graphicx}
\usepackage{amsbsy}

\usepackage{upgreek}
\usepackage{amsfonts}

\usepackage[usenames, dvipsnames]{color}

\usepackage{natbib}\biboptions{authoryear}

\begin {document}

\begin{frontmatter}

\title{Fast and accurate Voronoi density gridding from Lagrangian hydrodynamics data}

\author[mymainaddress]{Maya A. Petkova\corref{mycorrespondingauthor}}
\ead{map32@st-andrews.ac.uk}
\author[secaddress]{Guillaume Laibe}
\author[mymainaddress]{Ian A. Bonnell}
\cortext[mycorrespondingauthor]{Corresponding author}

\address[mymainaddress]{School of Physics and Astronomy, University of St Andrews, The North Haugh, St Andrews, KY16 9SS, UK}
\address[secaddress]{Univ Lyon, ENS de Lyon, CNRS, Centre de Recherche Astrophysique de Lyon UMR5574, F-69230, Saint-Genis-Laval, France}

\date{\today}

\begin{abstract}
Voronoi grids have been successfully used to represent density structures of gas in astronomical hydrodynamics simulations. While some codes are explicitly built around using a Voronoi grid, others, such as Smoothed Particle Hydrodynamics (SPH), use particle-based representations and can benefit from constructing a Voronoi grid for post-processing their output. So far, calculating the density of each Voronoi cell from SPH data has been done numerically, which is both slow and potentially inaccurate. This paper proposes an alternative analytic method, which is fast and accurate. We derive an expression for the integral of a cubic spline kernel over the volume of a Voronoi cell and link it to the density of the cell. Mass conservation is ensured rigorously by the procedure. The method can be applied more broadly to integrate a spherically symmetric polynomial function over the volume of a random polyhedron.
\end{abstract}


\end{frontmatter}


\section{Introduction}



Environments of astronomical interest and significance, such as stellar systems (\citet{Bate2002}, \citet{Clementel2014}), star-forming clouds (\citet{Dale2012}, \citet{Hubber2016}) or entire galaxies (\citet{Aumer2013}, \citet{Sijacki2012}), have been successfully simulated using computer models. While some of these models focus on the evolution of the fluid-like interstellar matter, others trace the stellar light that is propagating through this dusty, inhomogeneous medium in order to reproduce observations. The former ones are hydrodynamics computer codes, which can employ grid-based (Eulerian) or particle-based (Lagrangian) representations, and the latter are radiative transfer methods
.

Voronoi tessellations arise in astronomical computer modelling when a hybrid between grid-based and particle-based approaches is considered. A Voronoi tessellation is built around a set of generating sites (typically taken to be particle positions in this context), and each generating site is assigned the area of space that is closer to itself than to any other site (\citet{Voronoi1908}, \citet{Dirichlet1850}). This creates a grid structure consisting of randomly shaped convex polyhedra, and as such it provides a discretisation of space without the rigidity of a Cartesian grid. Some hydrodynamics codes are built explicitly around a moving Voronoi grid (\citet{Springel2010}, \citet{DuffellMacFadyen2011}, \citet{Vandenbroucke2016}) for the above reasons. Others, such as Smoothed Particle Hydrodynamics (SPH) (\citet{Lucy1977}, \citet{Gingold1977}), follow a particle representation, however their output is often post-processed with a radiative transfer code, requiring a density grid (\citet{Koepferl2016}, \citet{Clementel2014}, \citet{Hubber2016}). Due to the fact that SPH particles follow the fluid motion and are thus potentially irregular, the density profile is typically reconstructed using an adaptive mesh refinement scheme (\citet{Kurosawa2001}, \citet{Steinacker2002}, \citet{Harries2004}) or, more recently, a Voronoi grid (\citet{Camps2013}).


The problem addressed by this paper is how to calculate the average density of a Voronoi cell when regridding SPH data. Such task is non-trivial, as in SPH the density at every point in space is given as a sum of particle contributions (\citet{Price2012}). The contribution of each particle to the local density is calculated from the distance to the particle and the kernel function. The latter is a bell shaped compactly supported function that goes to zero at a finite radius, defined by the smoothing length. The SPH continuous density distribution constructed through the kernel function is an approximate representation of the true underlying density distribution, discretised by the SPH particles. Discussing the choice of kernel function and the errors associated with it is beyond the scope of this paper (see \citet{Price2012} and \citet{Monaghan1985} for more information). Instead from now on we will assume that the SPH kernel density distribution is a close representation of the true distribution, and as such we want to map it correctly onto a Voronoi grid.

One possible approach for calculating the density structure of the grid would be to divide the mass of each particle by the volume of its corresponding cell (\citet{Clementel2014}, \citet{Hubber2016}). This is only possible when there is one grid cell per particle, which is often the case as SPH particle positions are commonly used as grid-generating sites for the Voronoi tesselation (\citet{Clementel2014}, \citet{Hubber2016}).  This method has been successfully adopted by some authors, since it is easy to implement and it ensures mass conservation between the particle and grid representations. However, it has two major limitations. The first one is related to the fact that there is no direct correspondence between the size of a Voronoi cell and the smoothing length of an SPH kernel, which can lead to an assigned cell density that deviates significantly from the local SPH density in that region. In particular, there is an unfortunate resolution effect when a large density gradient is present, resulting in some cells having too high and others too low densities. Such resolution issues can be overcome by inserting extra Voronoi cells (\citet{Koepferl2016}), which brings us to the second limitation of the method. It can only be applied when there is one grid cell per particle.

Koepferl et al. have implemented an alternative approach, in which they sample random points inside of each grid cell, compute the SPH density at each point and then calculate the average of these values (\citet{Koepferl2016}). Their method assigns accurate densities to the cells, however the thorough sampling of points can be a computationally slow process. Short computing times become crucial if we want to not only post process SPH data with radiative transfer approach, but also combine the two codes and run them together. In that case the Voronoi grid mapping will be performed many times, and optimising computing speed and mass conservation becomes compulsory.

This work offers an analytical solution to the above stated problem instead. We compute the mass contribution to a cell from each neighbouring SPH particle using a derived mathematical formula, add up these contributions to calculate the total mass contained in the cell, and then divide this mass by the volume of the cell in order to obtain the average density \footnote{\label{code-link}An implementation of the code can be downloaded from https://github.com/mapetkova/kernel-integration.}. We demonstrate the validity of our method in Section \ref{sectionMethod} and show the detailed mathematical derivations in Section \ref{sectionDerivation}. In the remaining sections we address the testing of the method in terms of computing time and accuracy, and we compare its results obtained for specific datasets with those from other approaches used in the field.



\section{Method} \label{sectionMethod}
\subsection{SPH kernel function and density estimation}
One of the most important questions in the heart of SPH is how to obtain a continuous density profile from a set of discrete mass particles (\citet{Price2012}). In order to have a measure of density at a given point in space, one needs to consider the local particle distribution. Furthermore, particles that are further away from the site of interest should have a lesser contribution than ones closer to that site. Arising from these considerations, the density at point $\mathbf{r}$ in SPH is calculated using the following expression:

\begin{equation}
\rho(\mathbf{r}) = \sum_{j=1}^N m_j W(|\mathbf{r}-\mathbf{r}_j|,h_j).
\label{eqn1}
\end{equation}

In the above $m_j$ is the mass of a particle located at $\mathbf{r}_j$ and $W$ is the SPH kernel function, which depends on the distance between the particle and the point of interest. The parameter $h_j$ is called kernel smoothing length and it gives a measure of the "radius of influence" that a particle has towards the local density around it. 

$W$ is normalised to ensure that the total mass of a region described by the continuous density distribution equals the sum of particle masses, as it should. 

\begin{equation}
M = \int_{V}  \rho(\mathbf{r'}) \mathrm{d}V' = \sum_{j=1}^N m_j,
\label{eqn2}
\end{equation}

where $V$ is the volume of the entire region of space in the simulation. From equations \ref{eqn1} and \ref{eqn2}, this requires:

\begin{equation}
\int_{V}  W(|\mathbf{r'}-\mathbf{r}_j|,h_j) \mathrm{d}V' = 1.
\end{equation}

\subsection{Voronoi cell density}
We will now move from the concept of density at a specific location to that of average density in an enclosed volume (e.g. a Voronoi cell). A Voronoi tessellation divides space in $N'$ non-overlapping regions (note that $N'$ can be chosen to differ from the total number of particles, $N$), such that all of them add up to the full volume of $V$. Let us denote the average density of the $i$-th region by $\rho_i$.

\begin{equation}
\rho_i \equiv \langle \rho \rangle = \frac{1}{V_i} \int_{V_i}  \rho(\mathbf{r'}) \mathrm{d}V',
\end{equation}

where $V_i$ is the region's volume.

The above expression can also be written in terms of $M_i$, the total mass contained in the region:

\begin{equation}
\rho_i = \frac{M_i}{V_i} ,
\end{equation}

with

\begin{eqnarray}
M_i & =  & \int_{V_i}  \rho(\mathbf{r'}) \mathrm{d}V' \\
 & = & \int_{V_i}  \sum_{j=1}^N m_j W(|\mathbf{r'}-\mathbf{r}_j|,h_j) \mathrm{d}V'\\
 & = & \sum_{j=1}^N m_j \int_{V_i}  W(|\mathbf{r'}-\mathbf{r}_j|,h_j) \mathrm{d}V'.
\end{eqnarray}

The last representation of $M_i$ allows us to think of a "mass contribution" that each particle has towards a region. This contribution is given by the integral of the kernel function over the volume of the region.

Since SPH typically uses kernel functions that have an analytic description, one could, in principal, attempt to solve the integral of interest analytically. The derivation of one such solution is the focus of the following section.

\subsection{Mass conservation}

As a final step, let us demonstrate that the masses of the regions, as considered above, add up to the sum of particle masses. We can write the total mass as:

\begin{eqnarray}
\sum_{i=1}^{N'} M_i & =  & \sum_{i=1}^{N'} \sum_{j=1}^N m_j \int_{V_i}  W(|\mathbf{r'}-\mathbf{r}_j|,h_j) \mathrm{d}V'\\
& = &  \sum_{j=1}^N m_j \sum_{i=1}^{N'} \int_{V_i}  W(|\mathbf{r'}-\mathbf{r}_j|,h_j) \mathrm{d}V'.
\end{eqnarray}


Since the different Voronoi cell regions are non-overlapping and together they cover the full simulation space, we can rewrite:

\begin{equation}
\sum_{i=1}^{N'} \int_{V_i}  W(|\mathbf{r'}-\mathbf{r}_j|,h_j) \mathrm{d}V' = \int_{V}  W(|\mathbf{r'}-\mathbf{r}_j|,h_j) \mathrm{d}V' = 1,
\end{equation}

using the normalisation property of the kernel.

This gives us $\sum_{i=1}^{N'} M_i = \sum_{j=1}^N m_j$, as desired. Mass conservation is therefore ensured to high level of precision even in simulations where the mapping from SPH to a Voronoi grid is repeated many times (e.g. if the radiative transfer feedback is included during the SPH runtime).

\section{Derivation} \label{sectionDerivation}
\subsection{Overview}
In order to simplify the notation, let us drop the $i$ and $j$ indices from the previous equations and rewrite the mathematical problem in an easier-to-work-with form. It is sufficient to consider one single particle, positioned at the origin of a coordinate system, together with a random polyhedron (which may or may not contain the origin itself). We can assume that we know the coordinates of each vertex of the polyhedron, enclosing space $\mathcal{V}$, as well as the smoothing length of the particle's kernel function, $h$. Thus, we will compute the integral:

\begin{equation}
I_{V} = \int_{\mathrm{\mathcal{V}}}  W(r) \mathrm{d}V.
\label{eqnIv}
\end{equation}

\begin{figure}[h!]			
	\centering
	\includegraphics[width=0.49\textwidth]{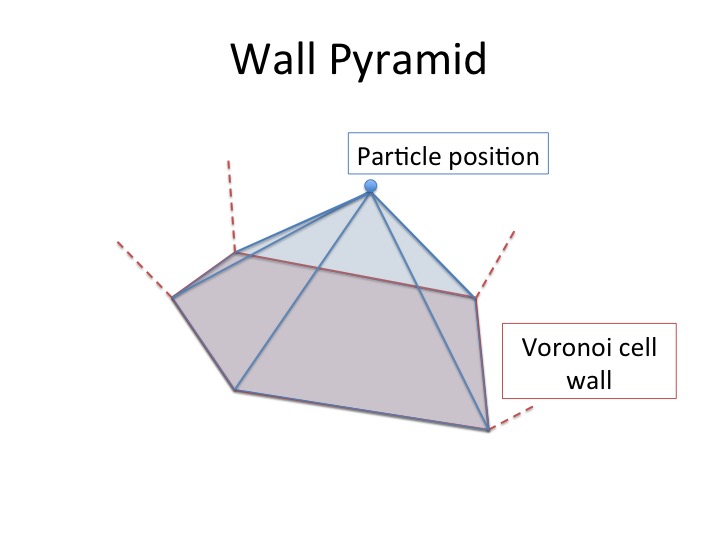}	
	\includegraphics[width=0.49\textwidth]{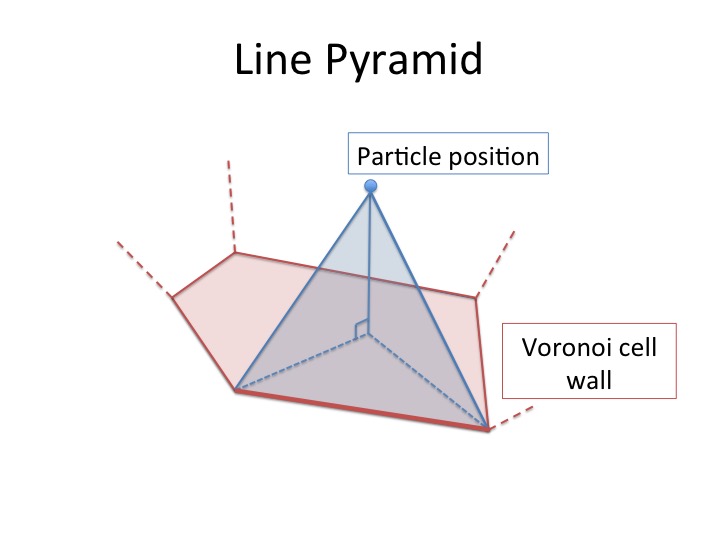}			
	\includegraphics[width=0.49\textwidth]{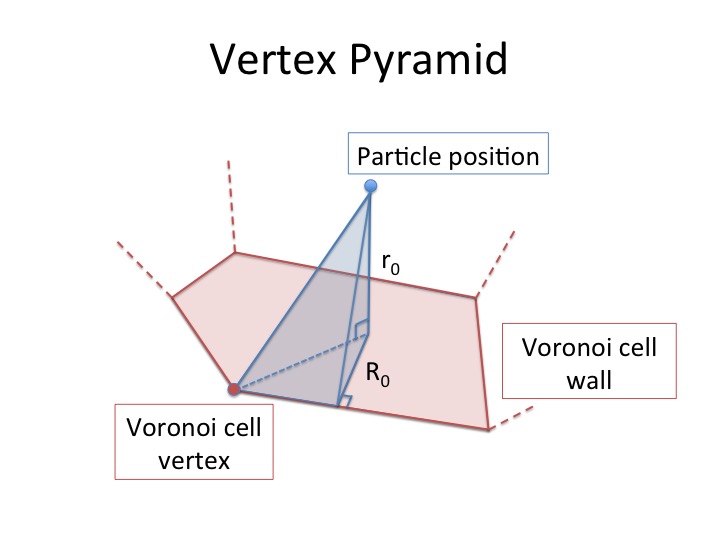}			
	
	\caption{By applying the Divergence Theorem (\textit{Top left}), a Voronoi cell is effectively divided into wall pyramids. When calculating the surface integral $I_S$ for each wall, we obtain the volume integral of the kernel function inside the shaded pyramid. By applying Green's Theorem (\textit{Top right}), a wall pyramid is divided into line pyramids. The line integral $I_L$ represents the volume integral of the kernel function inside the line pyramid. Furthermore, a line pyramid (\textit{Bottom}) comprises of two vertex pyramids. The expression for $I_P$, evaluated at each vertex of each wall, gives us the volume integral of the kernel function inside the vertex pyramid.}
	\label{derivation-figure}
\end{figure}

The solution presented in this section follows three logical steps. First, (step i) $I_V$ is transformed from a volume integral to a surface integral using the Divergence Theorem. The surface integral consists of a sum of 2D integrals (each of them denoted as $I_S$), calculated over the area of each of the polygonal faces of the polyhedron. Geometrically $I_S$ represents the volume integral of the kernel function inside of what we have named as a ``wall pyramid'' (see Figure \ref{derivation-figure}). Secondly, (step ii) each of the 2D integrals is then reduced to a contour integral along the edges of each wall using Green's Theorem. Similarly, the integral of each line segment ($I_L$) of the contour is associated with the integral of the kernel function inside of a ``line pyramid'' (Figure \ref{derivation-figure}). Finally, (step iii) one more integration is performed, so that in the end we have an expression for $I_V$ that only depends on the location of each vertex of the polyhedron. By evaluating our final solution at each vertex location, we obtain the integral of the kernel function inside the volume of a particular ``vertex pyramid'' (Figure \ref{derivation-figure}).

The mathematical steps of this derivation follow roughly the outline presented in \citet{Mirtich1996}. Some important modifications needed to be made, however, to accommodate for the spherically-symmetric nature of $W$ and the angular shape of the polyhedron.

\subsection{Kernel function}
For the purposes of this work we have picked a cubic spline kernel, given by the following expression:


\begin{equation}
   W(r) = \frac{1}{h^3\pi}  \begin{cases}
      1 - 1.5 \left( \frac{r}{h} \right )^2 + 0.75 \left ( \frac{r}{h} \right )^3, & r\leq h; \\
      0.25 \left (2 - \left (\frac{r}{h} \right ) \right )^3, & h\leq r\leq 2h; \\
      0, & r\geq 2h. \\
  \end{cases} \label{eq:torus}
\end{equation}

The final solution is kernel-specific, and this particular function was chosen because it is the most commonly used one in the astronomical community (\citet{Price2012}). Its polynomial form also makes it particularly suitable as it allows for simpler integration.

\subsection{Reducing the volume integral to a surface integral}
First, we will transform equation \ref{eqnIv} into a surface integral. In order to do so, we will use the Divergence Theorem, given by:

\begin{equation}
\int_{\mathcal{V}}  \nabla\cdot\mathbf{F} \mathrm{d}V = \int_{\partial \mathcal{V}}  \mathbf{F} \cdot \mathbf{\hat{n}} \mathrm{d}S,
\label{theoremDiv}
\end{equation} 

where $\hat{n}$ is the unit vector normal to the surface of the polyhedron $\partial \mathcal{V}$.

It is easy to notice that the left hand side of equation \ref{theoremDiv} is analogous to equation \ref{eqnIv}. In order to apply the theorem, we will construct $\mathbf{F}$, such that $\nabla\cdot\mathbf{F} = W$. This is analogous to the relationship between charge and electric field in electrostatics. 

Since $W$ is spherically-symmetric, expressing $\mathbf{F}$ in spherical coordinates is the most suitable choice. Hence for $\mathbf{F}$ and its divergence we have:

\begin{equation}
\mathbf{F} = F_{r} \mathbf{\hat{r}} + F_{\theta} \boldsymbol{\hat{\uptheta}} + F_{\phi} \boldsymbol{\hat{\upphi}};
\end{equation}

\begin{equation}
 \nabla\cdot\mathbf{F} = \frac{1}{r^2} \frac{\partial (r^2 F_{r})}{\partial r} + \frac{1}{r \sin \theta} \frac{\partial (F_{\theta} \sin \theta)}{\partial \theta} + \frac{1}{\sin \theta} \frac{\partial F_{\phi}}{\partial \phi}. 
\end{equation}

Additionally, due to the spherical symmetry of the problem, we consider a function for which $F_{\theta} = F_{\phi} = 0$. This gives us the equation

\begin{equation}
 \frac{1}{r^2} \frac{\partial (r^2 F_{r})}{\partial r} = W(r),
 \end{equation}

which integrates to the following solution:

\begin{equation}
   F_r(r) = \frac{1}{r^2} \frac{1}{h^3\pi}  \begin{cases}
      \frac{1}{3} r^3 - \frac{3}{10 h^2} r^5 + \frac{1}{8 h^3} r^6 + C_1, & r\leq h; \\
       & \\
      \frac{1}{4} \left (\frac{8}{3} r^3 - \frac{3}{h} r^4 + \frac{6}{5 h^2} r^5 - \frac{1}{6 h^3} r^6 \right ) + C_2, & h\leq r\leq 2h; \\
       & \\
      0 + C_3, & r\geq 2h, \\
  \end{cases} \label{eq:torus}
\end{equation}

where $C_1$, $C_2$ and $C_3$ are constants of integration. 

By enforcing continuity of $F_r$ at $r=h$ and $r=2h$, we obtain the relationships:

\begin{eqnarray}
C_1 = \frac{h^3}{60} + C_2; & & C_3 = \frac{4h^3}{15} + C_2.
\label{eqnC123}
\end{eqnarray}

Using the property of the kernel, that it integrates to $1$ within a radius of $r\geq 2h$, we can write that:

\begin{equation}
\int_{\partial \mathcal{V}}  \mathbf{F} \cdot \mathbf{\hat{n}} \mathrm{d}S = \int_{\phi=0}^{2\pi} \int_{\theta=0}^{\pi}  F_r(r) r^2 \sin \theta \mathrm{d} \theta \mathrm{d}\phi = 1.
\label{eqnFnorm}
\end{equation}

Equations \ref{eqnC123} and \ref{eqnFnorm} lead to the following solution for the constants of integration:

\begin{eqnarray}
C_1 = 0; & C_2 = -\frac{h^3}{60}; & C_3 = \frac{h^3}{4}.
\end{eqnarray}

\subsection{Calculating the surface integral on a plane}
We have now transformed $I_V$ into the surface integral given by equation \ref{theoremDiv}. Since the surface of a polyhedron is a set of polygons, we will focus on integrating over only one of them. Namely, we wish to compute the integral

\begin{equation}
I_S = \int_{\mathcal{A}}  \mathbf{F} \cdot \mathbf{\hat{n}} \mathrm{d}S,
\label{eqnIs}
\end{equation} 

where $\mathcal{A}$ is the region of space contained in a single polygonal wall.

Without loss of generality, we can choose the spherical coordinate system that $\mathbf{F}$ is defined in so that the normal of the plane of the desired flat surface coincides with the $\theta=0$ axis. Let the orthogonal distance from the origin to the plane be denoted by $r_0$.

We can express any surface element in spherical coordinates as 

\begin{equation}
\mathrm{d}\mathbf{S} = \mathbf{\hat{n}} \mathrm{d}S = r^2 \sin\theta \mathrm{d}\theta \mathrm{d}\phi \mathbf{\hat{r}} + r \sin\theta \mathrm{d}r \mathrm{d}\phi \boldsymbol{\hat{\uptheta}} + r \mathrm{d}r \mathrm{d}\theta \boldsymbol{\hat{\upphi}},
\end{equation}

and hence:

\begin{equation}
\mathbf{F} \cdot \mathbf{\hat{n}} \mathrm{d}S = F_r r^2 \sin\theta \mathrm{d}\theta \mathrm{d}\phi. 
\end{equation}

Since we are integrating over $\theta$ and $F_r$ is a function of $r(\theta)$, it would be necessary to pick a suitable integration variable. From the chosen coordinate system orientation we have the relationship:

\begin{equation}
r = \frac{r_0}{\cos\theta}.
\end{equation}

Let $\mu = \cos \theta$, then $r = \frac{r_0}{\mu}$ and $\mathrm{d}\mu = - \sin \theta \mathrm{d}\theta$. This gives us the following expressions, which are simpler to work with:

\begin{equation}
\mathbf{F} \cdot \mathbf{\hat{n}} \mathrm{d}S = - F_r(\mu) \left (\frac{\mu}{r_0} \right )^{-2} \mathrm{d}\mu \mathrm{d}\phi;
\end{equation}

\begin{equation}
   F_r(\mu) = \left ( \frac{\mu}{r_0}\right )^2 \frac{1}{h^3\pi}  \begin{cases}
      \frac{1}{3} \left (\frac{\mu}{r_0} \right )^{-3} - \frac{3}{10 h^2} \left (\frac{\mu}{r_0} \right )^{-5} + \frac{1}{8 h^3} \left (\frac{\mu}{r_0} \right )^{-6},  & \mu\geq \frac{r_0}{h}; \\
       & \\
      \frac{1}{4} \left (\frac{8}{3} \left (\frac{\mu}{r_0} \right )^{-3} - \frac{3}{h} \left (\frac{\mu}{r_0} \right )^{-4} + \frac{6}{5 h^2} \left (\frac{\mu}{r_0} \right )^{-5} - \frac{1}{6 h^3} \left (\frac{\mu}{r_0} \right )^{-6} - \frac{h^3}{15} \right ), & \frac{r_0}{2h}\leq \mu\leq \frac{r_0}{h}; \\
       & \\
      \frac{h^3}{4}, & \mu\leq \frac{r_0}{2h}. \\
  \end{cases} \label{eq:torus}
\end{equation}

\subsection{Reducing the surface integral to a contour integral}
In order to represent the surface integral $I_S$ as a contour integral we use a method analogous to the one above. In two dimensions we have Green's Theorem, given by:

\begin{equation}
\int_{\mathcal{A}}  \nabla\cdot\mathbf{H} \mathrm{d}A = \int_{\partial \mathcal{A}}  \mathbf{H} \cdot \mathbf{\hat{m}} \mathrm{d}l,
\label{greenthm}
\end{equation}

where $\mathbf{\hat{m}}$ is the unit vector normal to the contour $\partial \mathcal{A}$ of the area of integration $\mathcal{A}$.

In order to apply the theorem, we need to construct function $\mathbf{H}$, such that $\nabla\cdot\mathbf{H} \mathrm{d}A = \mathbf{F} \cdot \mathbf{\hat{n}} \mathrm{d}S$. While $\mathbf{F}$ is defined as a three-dimensional vector function, $\mathbf{H}$ should be in two dimensions, and the spherically-symmetric nature of $W$ suggests that we should define $\mathbf{H}$ in terms of polar coordinates:

\begin{equation}
\mathbf{H} = H_{R} \mathbf{\hat{R}} + H_{\phi} \boldsymbol{\hat{\upphi}};
\end{equation}

\begin{equation}
 \nabla\cdot\mathbf{H} = \frac{1}{R} \frac{\partial (R H_{R})}{\partial R} + \frac{1}{R} \frac{\partial H_{\phi} }{\partial \phi}.
\end{equation}

Let us set $H_{\phi}=0$. The area element can be written as $\mathrm{d}A =  R \mathrm{d}R \mathrm{d}\phi$, where $\phi$ is ensured to be the same as the three-dimensional coordinate used for $\mathbf{F}$ by aligning the coordinate systems' axes appropriately. This gives us the following expression:

\begin{equation}
\nabla\cdot\mathbf{H} \mathrm{d}A = \frac{\partial (R H_{R})}{\partial R} \mathrm{d}R \mathrm{d}\phi. \label{eqndivH}
\end{equation}

From geometrical considerations we can show that $R = r \sin\theta = r_0 \tan\theta$, which leads to $\mathrm{d}R = r_0 \sec^2\theta \mathrm{d}\theta$. This allows us to rewrite the following:

\begin{eqnarray}
 \mathbf{F} \cdot \mathbf{\hat{n}} \mathrm{d}S & = &F_r r^2 \sin \theta \mathrm{d}\theta \mathrm{d}\phi \\
  & = & F_r \frac{R^2}{\sin^2 \theta} \sin \theta \frac{\mathrm{d}R}{r_0 \sec^2 \theta} \mathrm{d}\phi \\
  & = & F_r \frac{R^2}{r_0 \tan^2 \theta} \sin \theta \mathrm{d}R\mathrm{d}\phi \\
  & = & F_r r_0 \sin \theta \mathrm{d}R\mathrm{d}\phi. \label{eqnFdotn}
\end{eqnarray}

By combining equations \ref{eqndivH} and \ref{eqnFdotn}, we obtain the integral:

\begin{equation}
H_R = \frac{1}{R} \int F_r r_0 \sin\theta \mathrm{d}R. 
\end{equation}

We will now write the above expression in terms of $\mu$, so that we can integrate it easily. We can notice that $\sin \theta \mathrm{d}R = r_0 \sec^2\theta \sin \theta \mathrm{d}\theta = - r_0 \mu^{-2} \mathrm{d} \mu$ , and with this modification the solution for $H_R$ becomes:

\begin{equation}
   H_RR = \frac{r_0^3}{h^3\pi}  \begin{cases}
      \frac{1}{6} \mu^{-2} - \frac{3}{40} (\frac{r_0}{h})^{2} \mu^{-4} + \frac{1}{40} (\frac{r_0}{h})^{3} \mu^{-5}  + \frac{B_1}{r_0^3}, & \mu\geq \frac{r_0}{h}; \\
       & \\
      \frac{1}{4} (\frac{4}{3} \mu^{-2} - (\frac{r_0}{h})\mu^{-3}+ \frac{3}{10} (\frac{r_0}{h})^{2} \mu^{-4} - \frac{1}{30} (\frac{r_0}{h})^{3} \mu^{-5} + \frac{1}{15} (\frac{r_0}{h})^{-3} \mu) + \frac{B_2}{r_0^3}, & \frac{r_0}{2h}\leq \mu\leq \frac{r_0}{h}; \\
       & \\
      - \frac{1}{4} (\frac{r_0}{h})^{-3} \mu + \frac{B_3}{r_0^3}, & \mu\leq \frac{r_0}{2h}. \\
  \end{cases} 
\end{equation}

In the above expression $B_1$, $B_2$ and $B_3$ are the constants of integration, which can be functions of $r_0$ and $h$.

\subsection{Deriving expressions for $B_1$, $B_2$ and $B_3$}
Consider integrating $\mathbf{F}$ over the area of a circle, extending from $\mu=1$ to $\mu=\mu_0$. Depending on the value of $r_0$ we would need to use different parts of the piecewise form of $\mathbf{F}$. This will result in the constants of integration $B_1$, $B_2$ and $B_3$ having different form depending on $r_0$, so that $\mathbf{H}$ gives answers consistent with those for $\mathbf{F}$.

In order to find expressions for them, let us start by considering $r_0\geq 2h$. This means that $\frac{r_0}{2h}\geq 1\geq \mu$, and we need to use only the third expression for $\mathbf{F}$, which gives us the following integral:

\begin{equation}
I_S = 2\pi \int_{\mu_0}^{1} F_r(\mu) \left (\frac{\mu}{r_0}\right )^{-2} \mathrm{d}\mu = \frac{1}{2} (1-\mu_0).
\end{equation}

If we were to apply Green's theorem and use $\mathbf{H}$, then the following should give us the same answer for all values of $\mu_0$:

\begin{equation}
I_S = \int  \mathbf{H} \cdot \mathbf{\hat{m}} \mathrm{d}l = \int_0^{2\pi} H_R R \mathrm{d}\phi = \frac{2B_3}{h^3} - \frac{1}{2} \mu_0.
\end{equation}

By comparing the coefficients of each term of the above polynomials, we get that $B_3 = \frac{h^3}{4}$.

Similarly, we then consider the case of $h \leq r_0 \leq 2h$. Here, we have that $\frac{r_0}{h} \geq 1 \geq \mu$, however, depending on the final integration value of $\mu_0$, we would either use the second polynomial of $\mathbf{F}$ or a sum of the second and the third one. By calculating the integral in two different ways, as shown previously, we can then obtain expressions for the constants. If $\mu_0 \geq \frac{r_0}{2h}$, then we get an expression for $B_2$, and if $\mu_0 \leq \frac{r_0}{2h}$, we can express $B_3$.

In the third case, when $r_0 \leq h$, we have three possibilities ($\mu_0 \geq \frac{r_0}{h}$; $\frac{r_0}{2h} \leq \mu_0 \leq \frac{r_0}{h}$; $\mu_0 \leq \frac{r_0}{2h}$), which give rise to expressions for $B_1$, $B_2$ and $B_3$ respectively.

The final polynomial forms of $B_1$, $B_2$ and $B_3$ are as follows:

\begin{equation}
B_1 = \frac{r_0^3}{4} \left ( -\frac{2}{3} + \frac{3}{10} \left ( \frac{r_0}{h} \right )^2 - \frac{1}{10} \left ( \frac{r_0}{h} \right )^3 \right );
\end{equation}

\begin{equation}
   B_2 = \frac{r_0^3}{4}  \begin{cases}
      -\frac{2}{3} + \frac{3}{10} \left ( \frac{r_0}{h} \right )^2 - \frac{1}{10} \left ( \frac{r_0}{h} \right )^3 - \frac{1}{5} \left ( \frac{r_0}{h} \right )^{-2}, & r_0\leq h; \\
       & \\
      -\frac{4}{3} + \left ( \frac{r_0}{h} \right ) - \frac{3}{10} \left ( \frac{r_0}{h} \right )^2 + \frac{1}{30} \left ( \frac{r_0}{h} \right )^3  - \frac{1}{15} \left ( \frac{r_0}{h} \right )^{-3}, & h\leq r_0\leq 2h; \\
  \end{cases} \label{eq:torus}
\end{equation}

\begin{equation}
   B_3 = \frac{r_0^3}{4}  \begin{cases}
      -\frac{2}{3} + \frac{3}{10} \left ( \frac{r_0}{h} \right )^2 - \frac{1}{10} \left ( \frac{r_0}{h} \right )^3 + \frac{7}{5} \left ( \frac{r_0}{h} \right )^{-2}, & r_0\leq h;  \\
       & \\
      -\frac{4}{3} + \left ( \frac{r_0}{h} \right ) - \frac{3}{10} \left ( \frac{r_0}{h} \right )^2 + \frac{1}{30} \left ( \frac{r_0}{h} \right )^3  - \frac{1}{15} \left ( \frac{r_0}{h} \right )^{-3} + \frac{8}{5} \left ( \frac{r_0}{h} \right )^{-2}, & h\leq r_0\leq 2h; \\
       & \\
      \left ( \frac{r_0}{h} \right )^{-3},  &  r_0\geq 2h. \\
  \end{cases} \label{eq:torus}
\end{equation}

\subsection{Calculating the contour integral on a line}
We have now reduced $I_S$ to a contour integral, which consists of a sum of line integrals (i.e. along the edges of the polygonal wall). Similarly to before, we will only consider the integral of $\mathbf{H}$ over a single line segment, $\mathcal{L}$. The integral that we will focus on is given by:

\begin{equation}
I_L = \int_{\mathcal{L}}  \mathbf{H} \cdot \mathbf{\hat{m}} \mathrm{d}l.
\end{equation}

Without loss of generality, we can select the orientation of the coordinate system such that the $\phi=0$ line is perpendicular to the line segment that we are interested in. Let the perpendicular distance to the line from the centre of the 2D polar coordinate system be denoted by $R_0$.

We then have the following expressions:

\begin{equation}
\mathbf{\hat{m}} = \mathbf{\hat{x}} = \cos \phi \mathbf{\hat{R}} - \sin \phi \boldsymbol{\hat{\upphi}}; 
\end{equation}

\begin{equation}
\mathbf{H} \cdot \mathbf{\hat{m}} = H_R \cos \phi.
\end{equation}

In order to express $\mathrm{dl}$, we will use its vector form given by:

\begin{equation}
\mathrm{d}\mathbf{l} = \mathrm{dR} \mathbf{\hat{R}} + R\mathrm{d}\upphi \boldsymbol{\hat{\upphi}}.
\end{equation}

Alternatively, we also have that:

\begin{equation}
\mathrm{d}\mathbf{l} = \mathrm{d}l \mathbf{\hat{y}} = \mathrm{d}l \sin\phi \mathbf{\hat{R}} + \mathrm{d}l \cos\phi \boldsymbol{\hat{\upphi}}.
\end{equation}

By comparing the $\boldsymbol{\hat{\upphi}}$ terms we can write that:

\begin{equation}
\mathrm{d}l = \frac{R \mathrm{d}\phi}{\cos \phi}.
\end{equation}

And hence, for a linear segment, we have that:

\begin{equation}
\mathbf{H} \cdot \mathbf{\hat{m}} \mathrm{d}l = H_R R \mathrm{d}\phi.
\end{equation}

\subsection{Calculating the line integral analytically}
In order to proceed, we require a kernel function which is integrable, in order to provide an analytical (or tabulated) form of this expression. Previously, we have expressed $H_R R$ as a function of $\mu$ and now we want to integrate it with respect to $\phi$. In order to complete the integration we need to express $\mu$ as a function of $\phi$ or vice versa. 

From geometrical considerations we have the following:

\begin{equation}
R = \frac{R_0}{\cos \phi} = r \sin \theta.
\end{equation}

And hence

\begin{equation}
r = \frac{R_0}{\sin \theta \cos \phi}.
\end{equation}

We also have that:

\begin{equation}
\mu = \frac{r_0}{r} = \frac{r_0 \sin \theta \cos \phi}{R_0}.
\end{equation}

By squaring both sides, substituting $\sin^2 \theta$ for $1-\mu^2$, and rearranging, we obtain the relationship:

\begin{equation}
\mu = \frac{\frac{r_0} {R_0} \cos \phi}{\sqrt{1+\frac{r_0^2} {R_0^2} \cos^2 \phi}}.
\end{equation}

Since $H_RR$ is a polynomial consisting of different powers of $\mu$, we need to integrate the following terms and insert them into the polynomial:

\begin{equation}
I_n = \int \mu^n \mathrm{d}\phi = \int \left ( \frac{\frac{r_0} {R_0} \cos \phi}{\sqrt{1+\frac{r_0^2} {R_0^2} \cos^2 \phi}} \right )^n \mathrm{d}\phi,
\label{eqnmuphi}
\end{equation}

where $n \in \mathbb{Z}$. 

We can easily notice that $I_0$ is trivial and can be expressed as:

\begin{equation}
I_0 = \int \mathrm{d}\phi = \phi + C.
\end{equation}

For the remaining even powers ($n=-2k$, $k \in \mathbb{N}$) of $\mu$ we can simplify as follows:

\begin{equation}
I_{-2k} = \int \left ( 1 + \frac{1}{\frac{r_0^2} {R_0^2} \cos^2 \phi} \right )^k \mathrm{d}\phi.
\end{equation}

Hence,

\begin{eqnarray}
I_{-2} &=& \phi + \int \frac{\mathrm{d}\phi}{\frac{r_0^2} {R_0^2} \cos^2 \phi} \\
          &=& \phi +  \frac{r_0^2} {R_0^2} \tan \phi + C;
\end{eqnarray}

\begin{eqnarray}
I_{-4} &=& \int \left ( 1 + \frac{2}{\frac{r_0^2} {R_0^2} \cos^2 \phi} + \frac{1}{\frac{r_0^4} {R_0^4} \cos^4 \phi}  \right ) \mathrm{d}\phi \\
          &=& \phi +  2\frac{r_0^2} {R_0^2} \tan \phi + \frac{1}{3}\frac{r_0^4} {R_0^4} \tan \phi (\sec^2 \phi + 2) + C.
\end{eqnarray}

For the odd powers ($n=1$; $n=-3$; $n=-5$) we will express $\phi$ in terms of $\mu$, as it follows from equation \ref{eqnmuphi}:

\begin{equation}
\mathrm{d}\phi = - \frac{R_0}{r_0} \frac{\mathrm{d}\mu}{(1-\mu^2) \sqrt{1-\left( 1 + \frac{R_0^2}{r_0^2} \right )\mu^2}}.
\end{equation}

Starting with integrating the expression for $n=1$, let $\alpha = \frac{R_0}{r_0}$, and then:

\begin{equation}
I_{1} = \int \frac{-\alpha \mu \mathrm{d}\mu}{(1-\mu^2) \sqrt{1-(1+\alpha^2) \mu^2}} 
\end{equation}

Let $u = \sqrt{1-(1+\alpha^2) \mu^2}$. Then $\mathrm{d}u = - \frac{(1+\alpha^2)\mu \mathrm{d}\mu}{\sqrt{1-(1+\alpha^2) \mu^2}}$, and $1-\mu^2 = \frac{\alpha^2 + u^2}{1+ \alpha^2}$. This changes the expression for $I_1$ to:

\begin{equation}
I_{1} = \int \frac{\alpha \mathrm{d}u}{\alpha^2 + u^2} = \tan^{-1} \left ( \frac{u}{\alpha} \right) + C.
\end{equation}

Using the same substitution, the expressions for $I_{-3}$ and $I_{-5}$ can be written as follows (for more details, see Appendix):

\begin{multline}
I_{-3} = \frac{\alpha(1+\alpha^2)}{4} \left ( \frac{2u}{1-u^2} + \log(1+u) - \log(1-u) \right ) \\ 
+ \frac{\alpha}{2} ( \log(1+u) - \log(1-u) ) + \tan^{-1} \left ( \frac{u}{\alpha} \right) + C;
\end{multline}

\begin{multline}
I_{-5} = \frac{\alpha(1+\alpha^2)^2}{16} \left ( \frac{10u-6u^3}{(1-u^2)^2} + 3(\log(1+u) - \log(1-u)) \right ) \\ 
+ \frac{\alpha(1+\alpha^2)}{4} \left ( \frac{2u}{1-u^2} + \log(1+u) - \log(1-u) \right ) \\ 
+ \frac{\alpha}{2} ( \log(1+u) - \log(1-u) ) + \tan^{-1} \left ( \frac{u}{\alpha} \right) + C.
\end{multline}

The final solution is hence given by:

\begin{equation}
   I_{P}= \frac{r_0^3}{h^3\pi}  \begin{cases}
      \frac{1}{6} I_{-2} - \frac{3}{40} \left (\frac{r_0}{h} \right )^2 I_{-4} + \frac{1}{40} \left (\frac{r_0}{h} \right )^3 I_{-5} + \frac{B_1}{r_0^3} I_0, & \frac{r_0}{h}\leq \mu; \\
       & \\
      \frac{1}{4} (\frac{4}{3} I_{-2} - (\frac{r_0}{h}) I_{-3}+ \frac{3}{10} (\frac{r_0}{h})^{2} I_{-4} - \frac{1}{30} (\frac{r_0}{h})^{3} I_{-5} + \frac{1}{15} (\frac{r_0}{h})^{-3} I_1) + \frac{B_2}{r_0^3} I_0, & \frac{r_0}{2h}\leq \mu\leq \frac{r_0}{h}; \\
       & \\
      - \frac{1}{4} (\frac{r_0}{h})^{-3} I_1 + \frac{B_3}{r_0^3} I_0, & \mu\leq \frac{r_0}{2h}. \\
  \end{cases} 
\end{equation}

In the applications of this method boundary conditions are applied to ensure continuity between the different regions of the function.

The method can be extended to any piecewise polynomial kernel using the integral relations in the Appendix.

\section{Application}
A similar derivation to the one above was also performed in 2D space (see Appendix). In that case, the step of applying the Divergence Theorem is omitted, since it is not relevant, and the starting kernel function has slightly different coefficients. The final solution, however, resembles the one obtained for the 3D case.

\subsection{Constructing a Voronoi grid}
There are two implementations of Voronoi tessellation construction algorithms that were used for this work. The first one is a two-dimensional one and is written in Fortran. It builds the tessellation by constructing its complementary Delaunay triangulation first, following an incremental algorithm (\citet{Bowyer1981}, \citet{Watson1981}).

The second one was used for the three-dimensional tests and applications, and was performed by the C++ library VORO++ (\citet{Rycroft2009}). The library is tailored towards performance efficiency, which is crucial for large-scale problems. It computes each Voronoi cell individually and stores statistical data, such as a list of neighbouring cells and cell volume. These features were beneficial for the implementation of our density calculation approach.

\subsection{Testing}



\begin{figure}[h!]			
	\centering
	\includegraphics[width=0.49\textwidth]{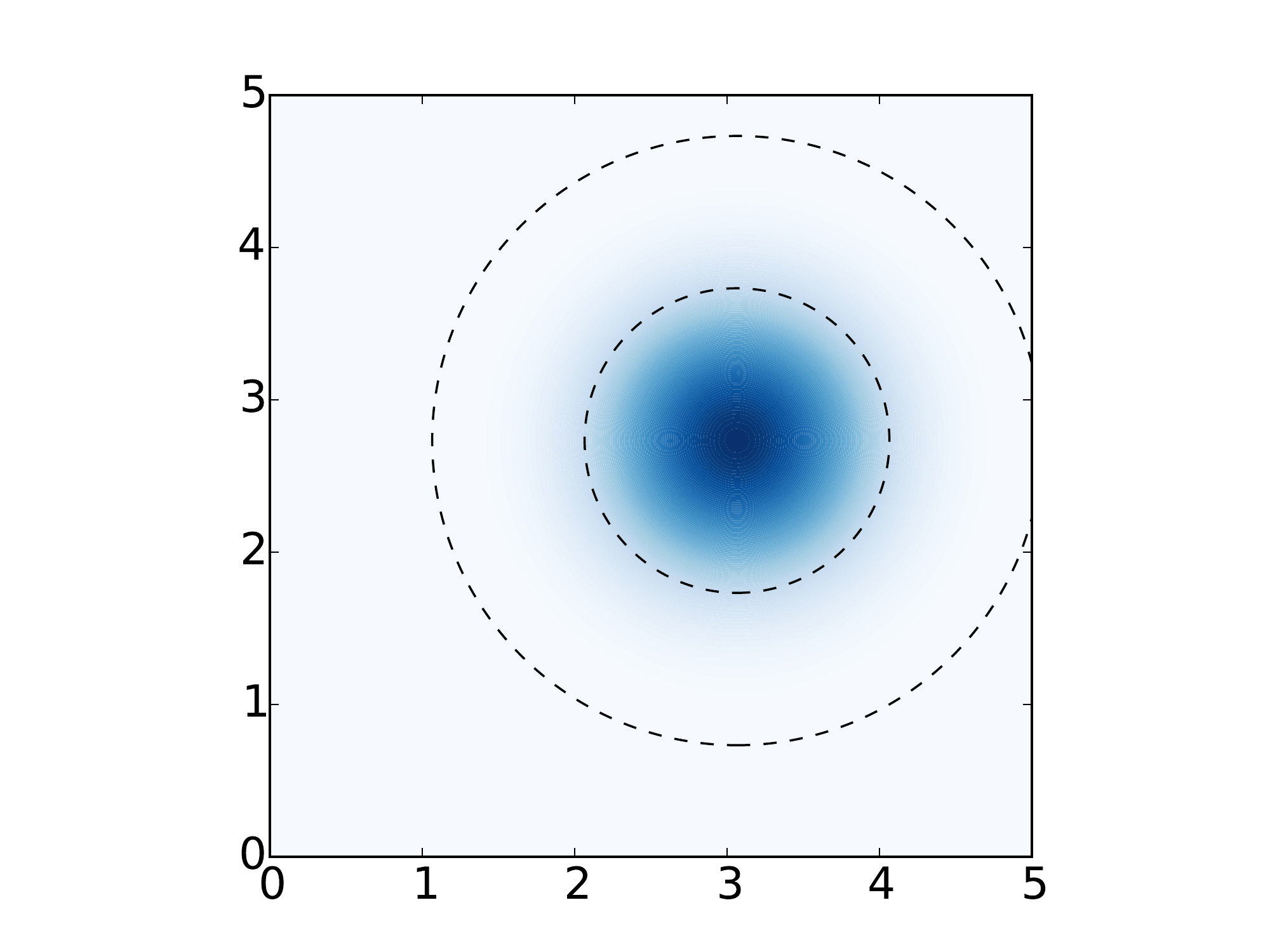}	
	\includegraphics[width=0.49\textwidth]{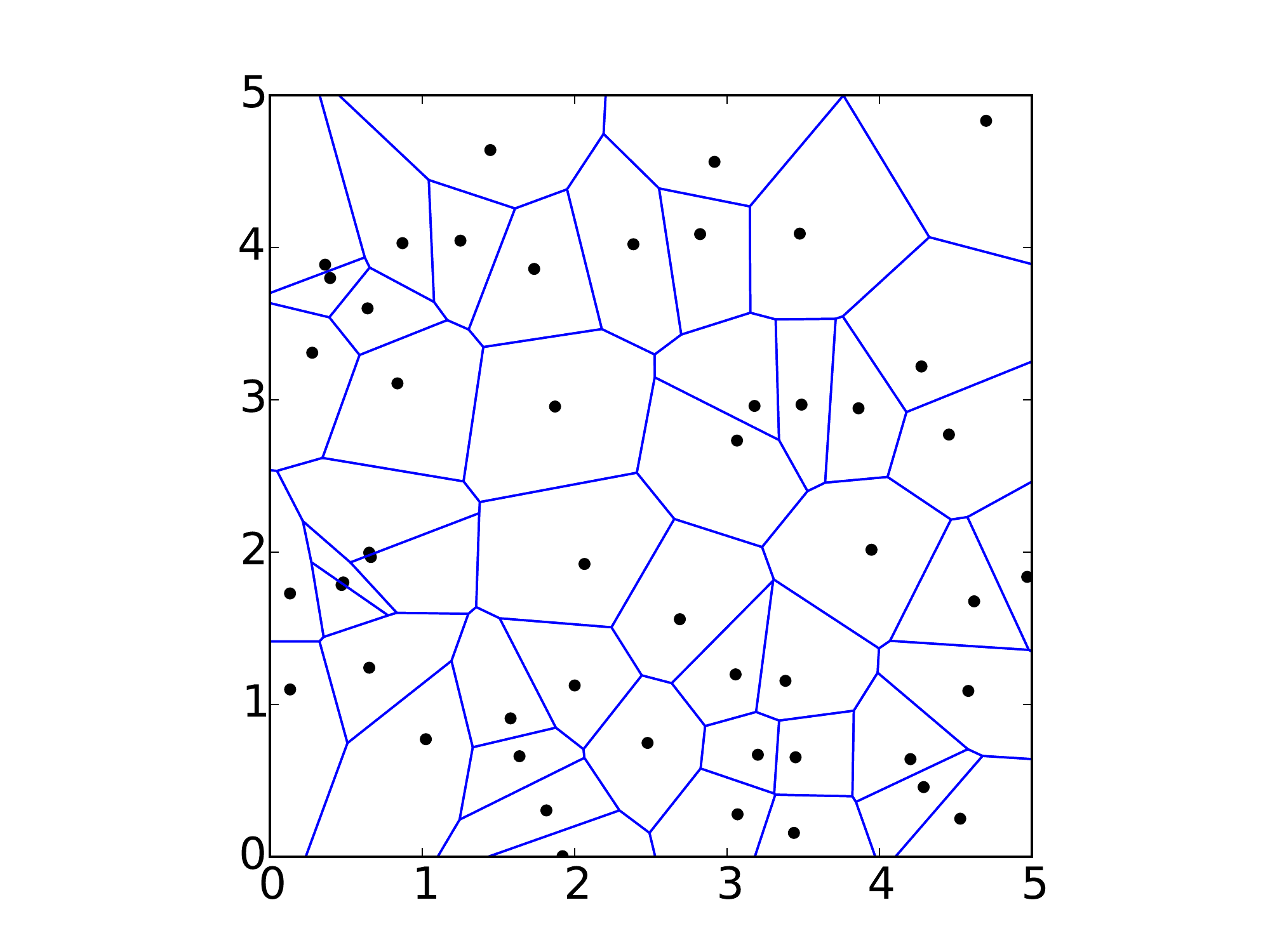}			
	\includegraphics[width=0.49\textwidth]{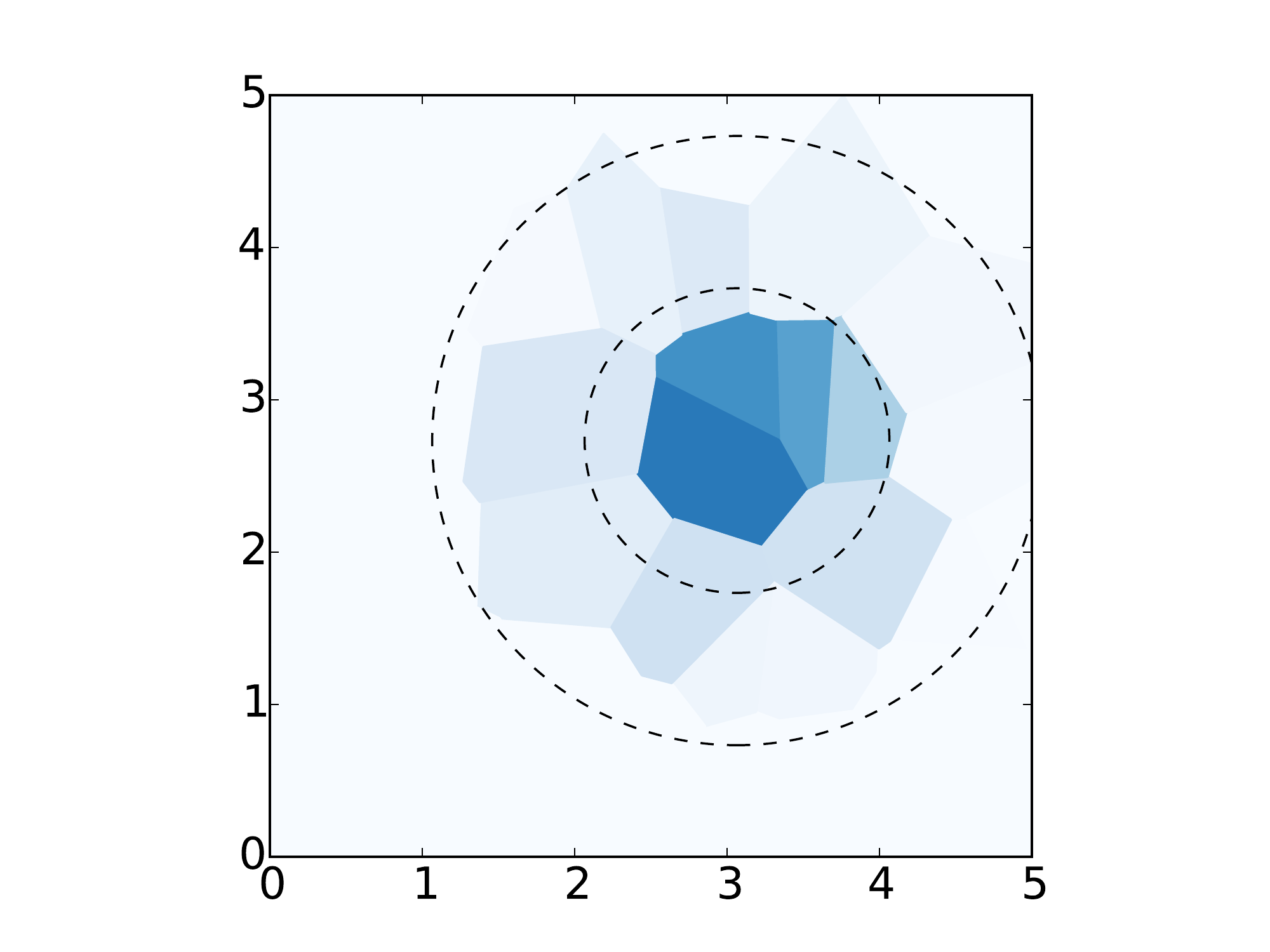}			
	
	\caption{(\textit{Top left}) 2D SPH kernel function (representing density) with $h=1$. Dark blue corresponds to higher density, and the contour lines are at $r=h$ and $r=2h$. The kernel function is zero outside of the larger cirlcle. (\textit{Top right}) 2D Voronoi grid created from 50 randomly sampled generating sites. (\textit{Bottom}) The average density of each Voronoi cell, as calculated with the analytic solution is plotted in colour. The same colour scheme is used as in the top left plot, and we can see that we preserve the SPH density structure.}
	\label{2d-1kernel}
\end{figure}

\begin{figure}[h!]			
	\centering
	\includegraphics[width=0.6\textwidth]{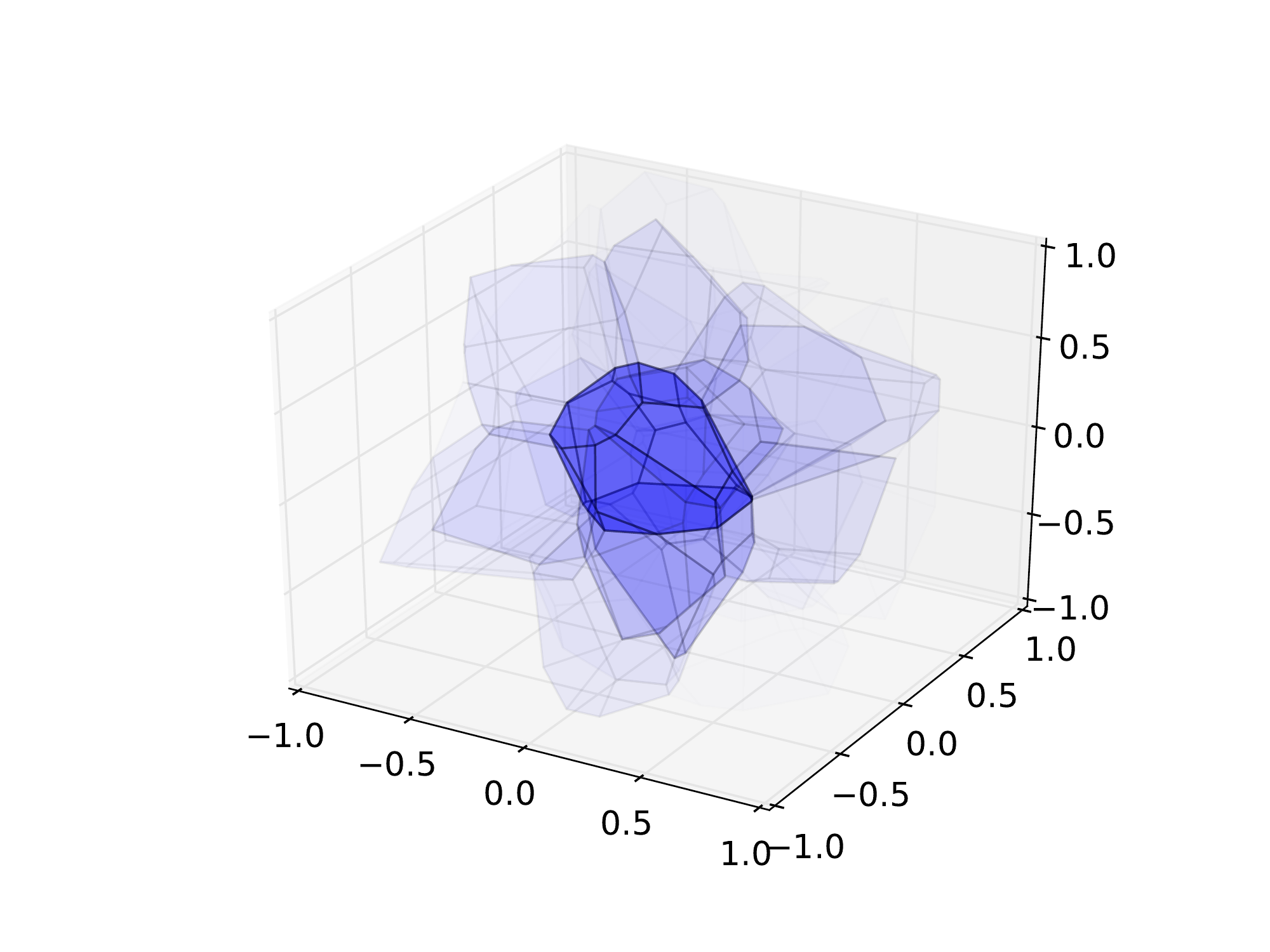}			
	\caption{Similar to Figure \ref{2d-1kernel}, but performed for a 3D test. A 3D SPH kernel function with $h=0.5$ is represented through a 3D Voronoi grid with 50 randomly sampled generating sites. The average density of each cell, as given by the analytic solution, is represented by its transparency. Darker regions correspond to higher column density along our line of sight. Here, again, the SPH density structure is preserved.}
	\label{3d-1kernel}
\end{figure}

Having completed the derivation of the analytic solution, the computer implementation follows a clear structure. For the case of a single particle existing in a Voronoi grid we loop through all of the cells and apply the mathematical formula for the ends of each side, of each wall, of a cell. We then add up these values obtained for each cell and divide the sum by the cell volume, which gives us the cell density. This setup served as our initial test and we performed it both in 2D and 3D (see Figure \ref{2d-1kernel} and \ref{3d-1kernel}).

\begin{figure}[h!]			
\centering
\includegraphics[width=0.65\textwidth]{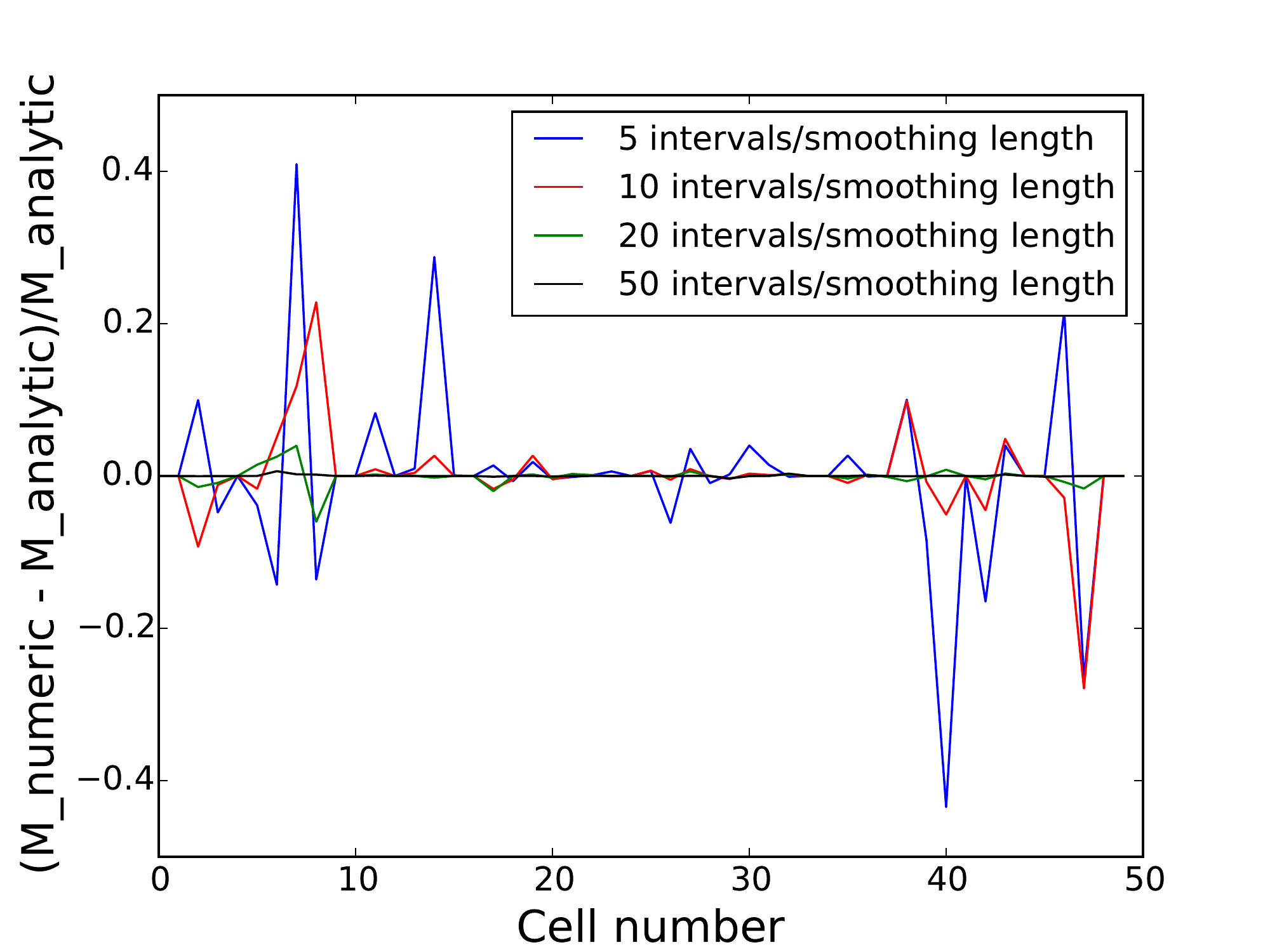}			
\caption{Fractional difference between the numerically calculated and the analytically calculated cell masses for the setup in Figure \ref{3d-1kernel}. The different colours represent the number of intervals per smoothing length that were used for the numerical integration of the cell masses. By increasing the number of intervals, the numerical masses converge to the analytically calculated ones, demonstrating the validity of our method.}
\label{residuals}
\end{figure}

Through simple visual inspection, the density distribution of the cells matches the expected profile of a particle's kernel. Furthermore, the total mass of all cells equals that of the particle that was considered. In order to have a more rigorous test, however, we have also implemented a numerical integration algorithm based on Simpson's rule and compared its answers for the cell masses to those given by the analytical solution (see Figure \ref{residuals}). The comparison shows clear agreement and demonstrates the validity of our proposed method.

\subsection{SPH data}
After completing the single kernel tests, we have applied the method to data taken from an SPH simulation of a clumpy cloud. The full data set contains 400728 particles and was produced with Phantom (Forgan and Bonnell (in prep.)) to simulate a clumpy shock. The dataset consists of high density cores embedded into a uniform low density medium, and was evolved for about 1700 years. For this example we have considered the last snapshot of the SPH run and have constructed a single grid cell around the position of each particle.

Once we have a large number of kernels and cells, it is no longer practical to loop over all cells for each kernel. Since the kernel function is zero beyond twice the smoothing length, it is sufficient to consider only cells that are within that radius. To do so, we employ a simple friends-of-friends type of algorithm. We construct a queue structure where we store the indices of the cells which will receive non-zero mass contribution from the particle. As we go through the queue, we check if each wall of the currently considered cell is within a radius of two smoothing lengths. If the entire wall or part of it is inside of this radius, we add the cell on the other side of the wall to the back of the queue, provided it is not already in it. Initially the queue starts with only one element, which is the index of the cell generated around the particle's location. 

We have considered subsets of the full SPH dataset in order to perform timing tests. In this specific case the number of cells that receive mass from a given particle is between about 50 and 100, hence for the larger subsets we can treat that number as constant and expect the computation time of the full density calculation to depend linearly on the number of particles in the subset.

This linear relation does not necessarily hold if we compare full SPH datasets of different total particle numbers. For the purposes of numerical convergence, an ideal implementation of SPH would have the number of neighbours scaling as the square root of the total particle number (\citet{Zhu2015}). This would lead to the computing time being proportional to the total number of particles to the power of 1.5 when different SPH datasets are compared to each other.

\begin{figure}[h!]			
\centering
\includegraphics[width=0.49\textwidth]{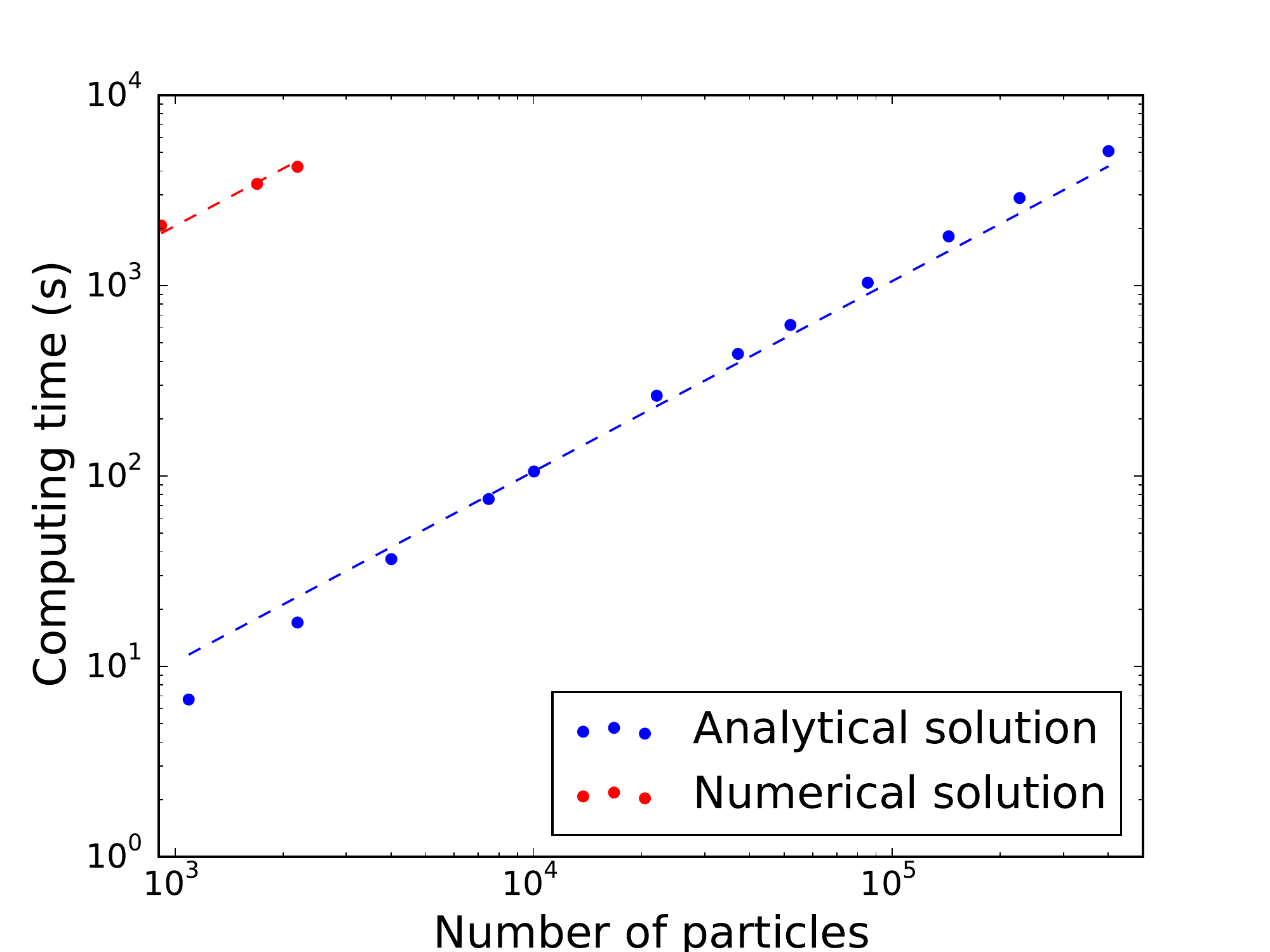}			
\caption{Comparison between the computing time of the analytically calculated density structure and the numerically calculated density structure (with 10 intervals per smoothing length) for differently sized samples of an SPH dataset. In both the analytic and the numeric cases, there is a linear dependence between computing time and the number of particles, as expected. The gradient of the numerical solution is 200 times larger than that of the analytic one, making the method much slower than desired.}
\label{timing}
\end{figure}

We can see in Figure \ref{timing} that the computing time for the analytical method follows a linear dependence, as expected. We have compared this to the numerical solution with 10 intervals per smoothing length, for which computational time also depends linearly on the number of particles. As the figure shows, the numerical approach is a factor of about 200 times slower than the analytical one. Note that both of the algorithms were executed sequentially, however they could be easily parallelised.

\section{Discussion}
\subsection{Comparing the analytically obtained densities to other types of density mapping}
\begin{figure}[h!]			
	\centering
	\includegraphics[width=0.49\textwidth]{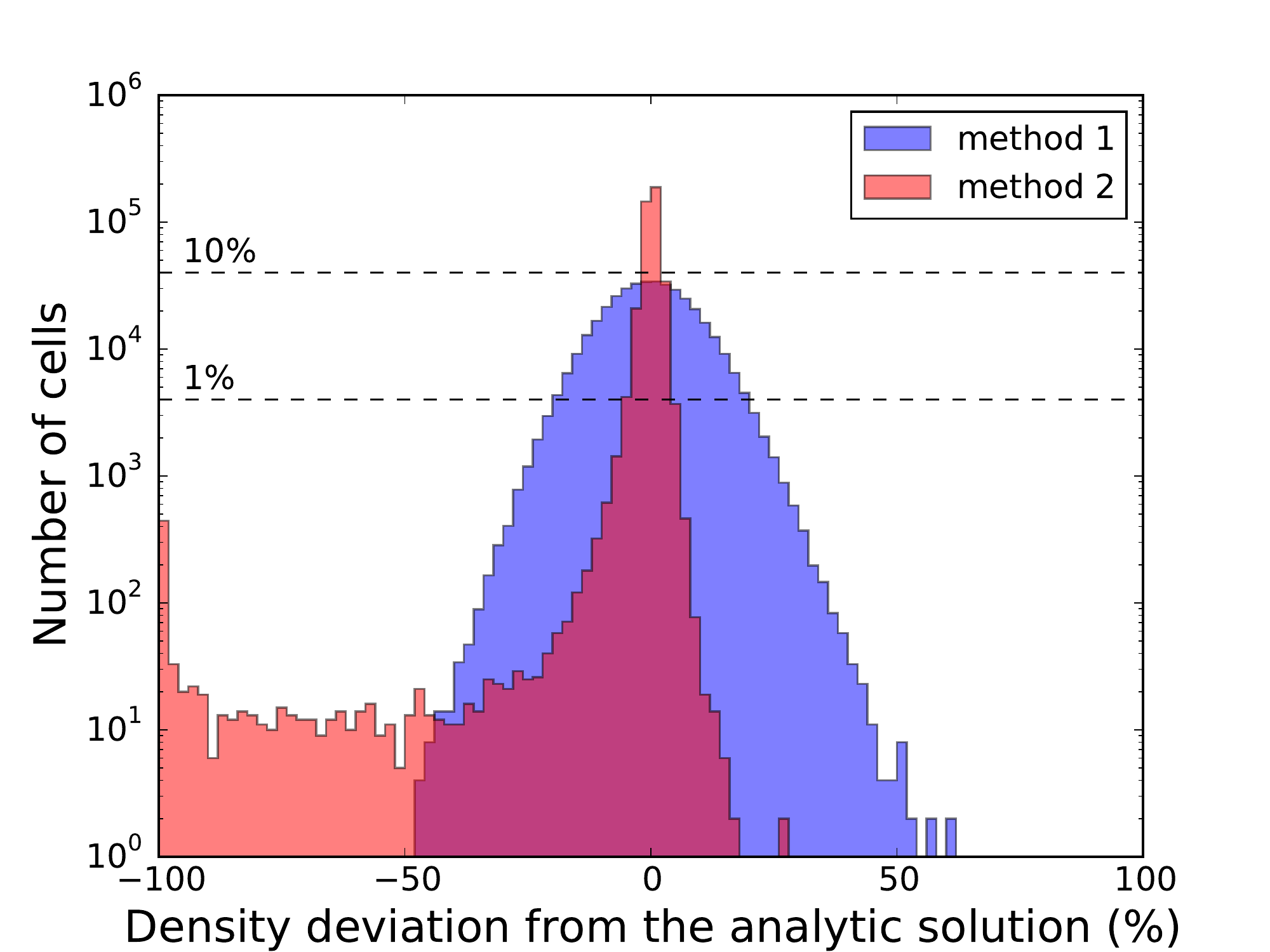}			
	\includegraphics[width=0.49\textwidth]{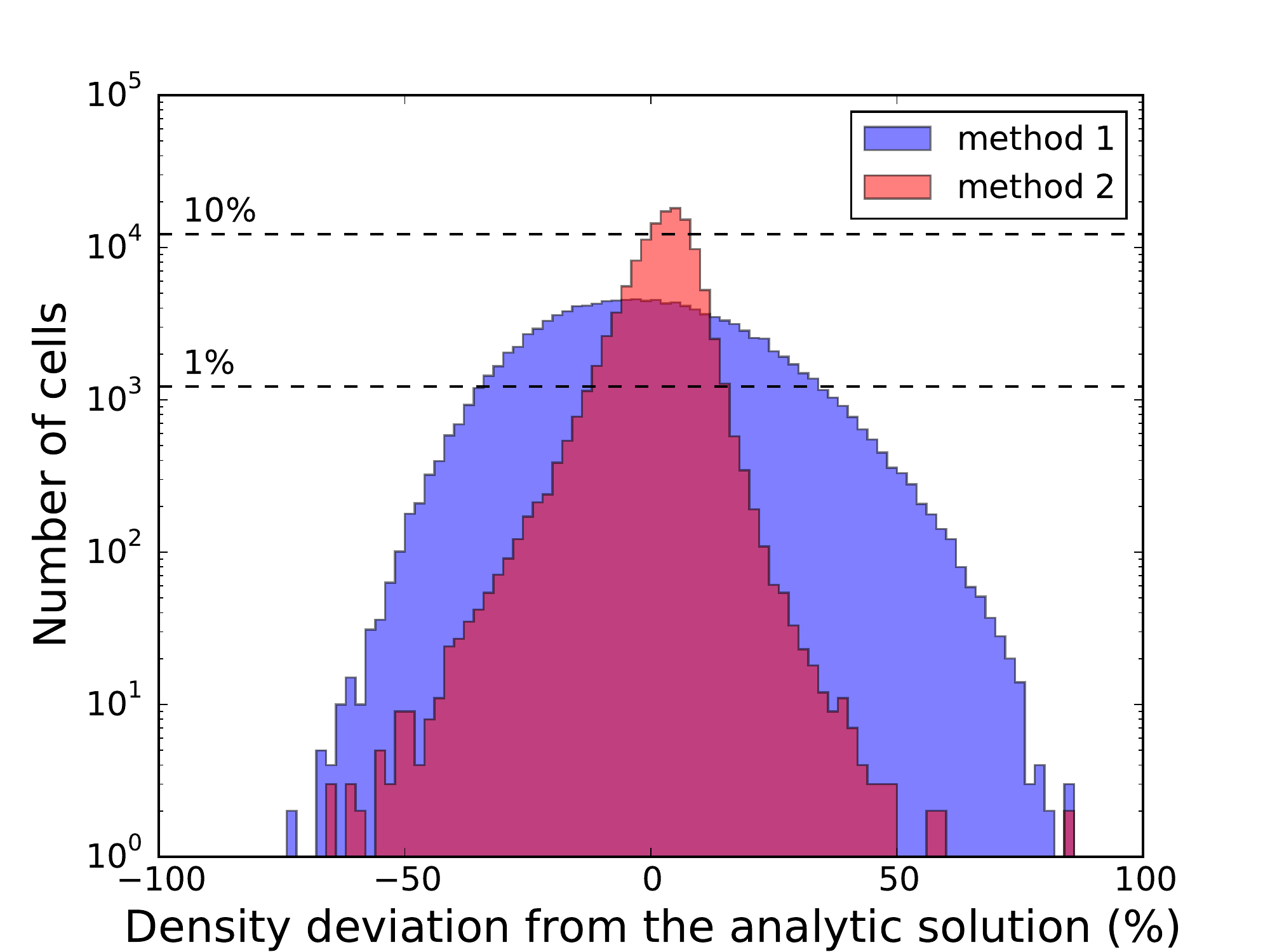}			
	\includegraphics[width=0.49\textwidth]{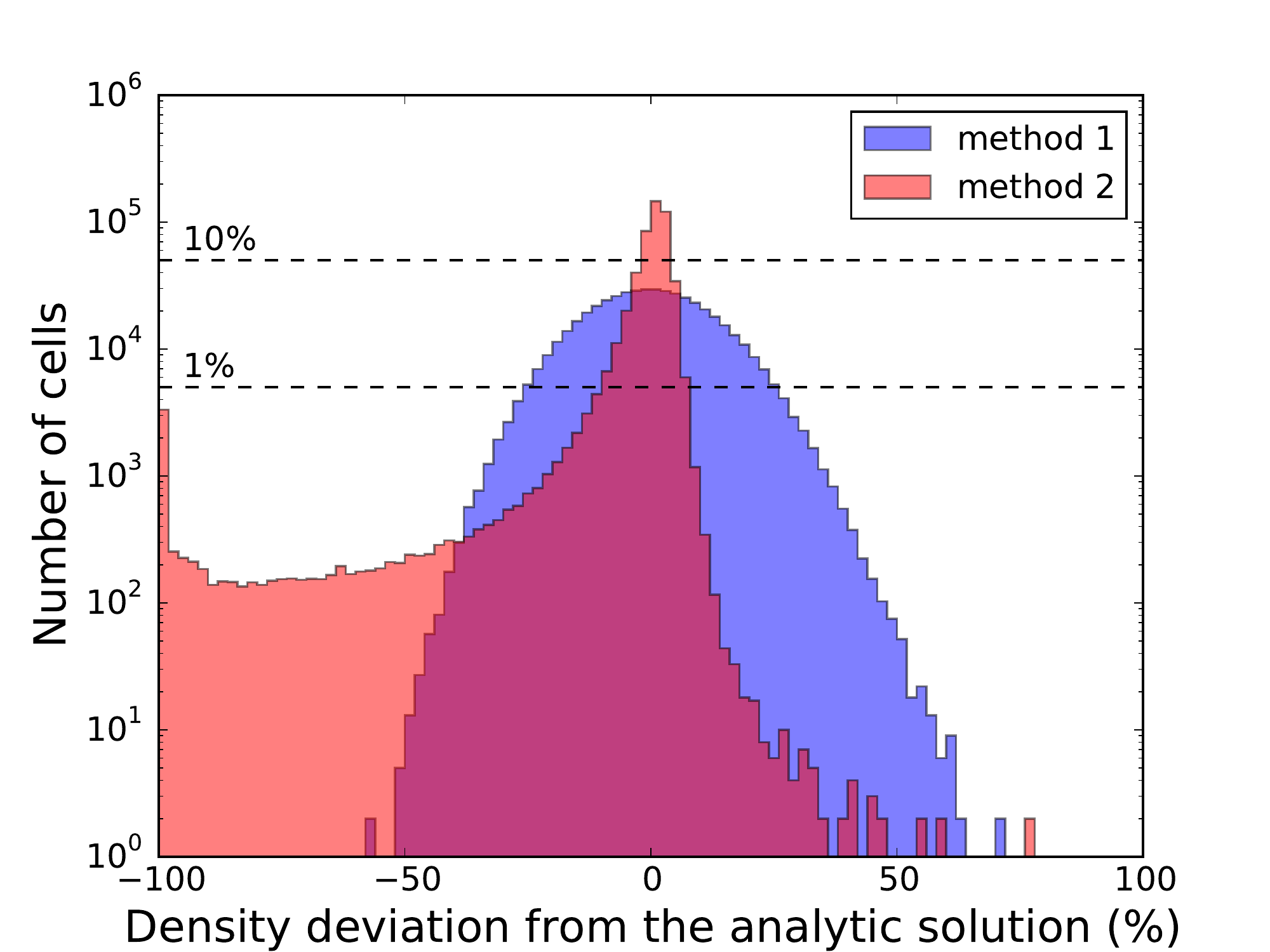}			
	\includegraphics[width=0.49\textwidth]{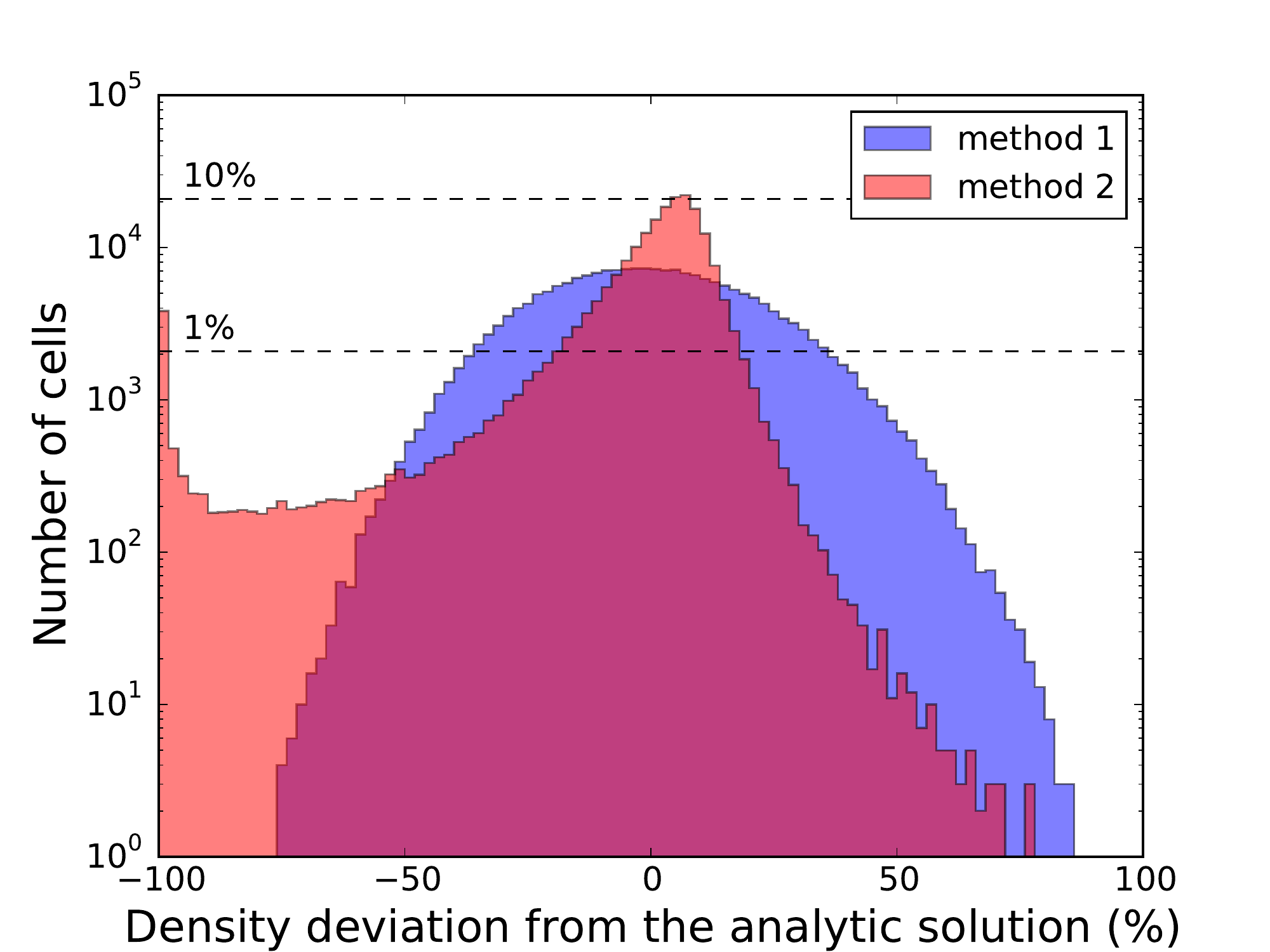}			
	\caption{Histograms studying the accuracy of two commonly used methods for density calculation: particle mass divided by cell volume (method 1) and SPH density at the cell centroid point (method 2). Method 1 and 2 are compared to the analytically obtained density for the case of a clumpy cloud (top left), uniform density box (top right), disk galaxy (bottom left) and supernova shock wall (bottom right). The dashed horizontal lines indicate the level at which the bins have reached 1\% and 10\% respectively of the total number of cells. Both method 1 and 2 show significant deviations from the analytic solution, which can cause inaccuracies when the SPH data is post-processed with MCRT.}
	\label{density_error}
\end{figure}

After demonstrating the validity of our proposed density calculation method, we will focus on comparing its results with the more commonly used ways of calculating average densities of Voronoi cells in astronomy. The first one, from here on referred to as method 1, is by creating one cell per particle and dividing a particle's mass by its cell volume to obtain density. We have previously argued that method 1 will not produce accurate solutions, and we have compared the results it gives to the analytical method in Figure \ref{density_error}. The second method, or method 2, calculates the SPH density at the centroid point of a Voronoi cell and assigns it as the average cell density. While method 2 can produce a reasonable estimate of the local density (see Figure \ref{density_error}) provided that no large density gradients are present in the data, it does not ensure mass conservation.

Figure \ref{density_error} demonstrates a comparison between method 1, method 2 and the analytic solution for four different datasets with varying properties, namely a clumpy cloud, a uniform density box, a disk galaxy and a cloud affected by a supernova explosion. The clumpy cloud is the dataset introduced in the previous section for testing purposes. The rest of the models were done using SPHNG (\citet{Bate1995}). The box consists of 122333 uniformly sampled particles, however some noise is present in the particle positions and hence in the densities. We have only used the initial setup of this model instead of evolving it in time, in order to keep the desired density distribution. We have done the same with the disk galaxy model (Ramon Fox et al. (in prep.)), by only considering its initial conditions given by the prescription of \citet{McMillan2007}. It contains 500000 gas particles, following a smooth density power-low without any features. Finally, the post-supernova cloud contains 208155 particles on both sides of a shock wall where many complex structures are present (Lucas et al. (in prep.)).

From Figure \ref{density_error} we can see that while method 2 tends to have broader range of density deviations than method 1, the bulk of its particles are concentrated in a narrower region around 0\%. Additionally to the primary peak method 2 has a secondary peak at -100\% for three of the four datasets. These are cells for which the centroid method assigns zero density, and are located at the outskirts of the data samples, where the cells are elongated due to the cuboid boundaries. The dataset without a secondary peak is the one of the uniform density cube, where the distribution of the data points matches the shape of the cuboid boundaries.

We have previously mentioned that method 2 does not ensure mass conservation. In each of the four examples that we have considered, the total mass produced by method 2 differs by less than 5\% from the analytic total mass.

The conversion from particle to grid structure of SPH data is typically done in order to perform post-processing with Monte Carlo Radiative Transport (MCRT), and it is hard to estimate how much the density deviation of both method 1 and method 2 will affect the results of the MCRT, since that will strongly depend on other parameters as well. However if we incorporate MCRT into the hydrodynamics code, we will be performing the density mapping and the radiative transport many times which will multiply the effect of any inaccuracies.

Although from Figure \ref{density_error}, method 2 appears preferable to method 1, especially when there are not many elongated cells at the boundaries, neither can provide a comparable solution to our analytical method.

\subsection{Allowing for the number of cells and the number of particles to differ}
A significant byproduct of our density calculation method is the fact that the number of cells does not need to match the number of particles. This was already evident in Figures \ref{2d-1kernel} and \ref{3d-1kernel}, where we presented the test cases. The implication of this byproduct is that we have gained extra freedom in choosing local grid resolution, making our method non-particle based. While it is a reasonable idea to construct a grid cell around each SPH particle, this will not be optimal if we are interested in post processing the data with MCRT.

Dense areas of a cloud, which are far away and/or shielded from the sources of light will receive very small number of photon packets during post-processing. Additionally, dense areas in SPH consist of a clustering of particles, which in this case will provide high resolution at a place where it is not needed. In terms of computing efficiency, this region will be much better represented by a single cell.

Conversely, at a sharp boundary between a high density and a low density region of a cloud, there is a poor resolution effect caused by the property of the Voronoi tessellation to bisect the distance between neighbouring cell generating sites. This creates elongated dense cells which stretch into the low density region, and can affect the MCRT post processing. This scenario has been studied by \citet{Koepferl2016} in the context of the geometry of an ionisation bubble, and their proposed solution is to create more cells within the region of interest.

\subsection{Broader applicability of the analytic solution}
In this paper, we have been considering a specific SPH kernel function in the form of a cubic spline. While this is one of the most broadly used SPH kernels, one might want to perform the density calculation method with a different function. The method presented in this work can be applied to other functions as well, provided that the individual integration steps can be performed. In particular, similar analytic solutions can be obtained for different polynomial kernels.

Additionally, this work is not limited to Voronoi cells. Throughout our derivations we have not used any of the geometric properties of the Voronoi tessellation, but have instead considered the space contained in a random polyhedron. This makes our method applicable to a broader range of problems, including calculating densities of regular grids.

Since the mathematical question that we have answered was about how to integrate a spherically symmetric function over the volume of any random polyhedron, the presented approach can have applications beyond SPH kernels and Voronoi cells as well.

\section{Summary}
We have presented a new approach for computing Voronoi cell densities from SPH data. Our method uses a cubic spline kernel function in order to calculate the mass contribution of a particle to the cells that it overlaps with. All of the mass contributions that a cell receives are then added up and divided by the cell volume in order to obtain the average density of the cell.

Our method is based on an analytic expression which we have derived both for 2D and 3D space, and which is evaluated at the vertices of each Voronoi polyhedron. When compared to a numerical solution for the same problem, our approach is about 200 times faster in terms of computational time.

We have additionally applied our solution to different SPH datasets and compared the density profiles that we have obtained to ones constructed with more commonly used density mapping methods. In some cases, we have found significant discrepancies, which can affect further post-processing of the SPH output.

A significant property of the presented work is that it is not limited to grid representations with equal number of SPH particles and Voronoi cells. This provides additional freedom to choose to alter the local resolution when post-processing SPH data sets.

Finally, the mathematical method of the derivation can be used more broadly. One could consider other kernels, or even functions unrelated to SPH or the current problem of interest, provided that the relevant integration steps can be performed. Moreover, the analytic integration was performed over the volume of a random polyhedron, which need not be a Voronoi cell. It can be applied to any grid representation with flat cell walls, including regular grids.

\section*{Acknowledgments}
The authors would like to thank Duncan Forgan, Felipe Ramon Fox and William Lucas for providing diverse datasets that this method could be tested on. Additionally, MAP would like to thank Duncan Forgan and Daniel Price for helpful discussions throughout the completion of this work.

MAP and IAB acknowledge funding from the European Research Council for the FP7 ERC advanced grant project ECOGAL. This work used the DiRAC Complexity system, operated by the University of Leicester IT Services, which forms part of the STFC DiRAC HPC Facility (www.dirac.ac.uk). This equipment is funded by BIS National E-Infrastructure capital grant ST/K000373/1 and STFC DiRAC Operations grant ST/K0003259/1. DiRAC is part of the National E-Infrastructure. GL acknowledges financial support from PNP, PNPS, PCMI of CNRS/INSU, CEA and CNES, France.

\section*{References}
\bibliography{kernel-integration}

\newpage
\section*{Appendix}
\subsection*{Derivation of $I_{-1}$}
In order to obtain $I_{-3}$ and $I_{-5}$, we will first find $I_{-1}$, as follows:

\begin{eqnarray}
I_{-1} &=& \int \frac{-\alpha \mathrm{d}\mu}{\mu (1-\mu^2) \sqrt{1-(1+\alpha^2) \mu^2}} \\ 
          &=& \int \frac{-\alpha (1-\mu^2 + \mu^2) \mathrm{d}\mu}{\mu (1-\mu^2) \sqrt{1-(1+\alpha^2) \mu^2}} \\
          &=& \int \frac{-\alpha \mathrm{d}\mu}{\mu \sqrt{1-(1+\alpha^2) \mu^2}} + \int \frac{-\alpha \mu \mathrm{d}\mu}{(1-\mu^2) \sqrt{1-(1+\alpha^2) \mu^2}} \\
          &=& I_{-1}^{'} + I_1.
\end{eqnarray}

As we already have a solution for $I_1$, we only need to express $I_{-1}^{'}$ by using $u = \sqrt{1-(1+\alpha^2) \mu^2}$ and the fact that $\mu = \frac{1-u^2}{1+\alpha^2}$:

\begin{eqnarray}
I_{-1}^{'} &=& \int \frac{\alpha \mathrm{d}u}{1-u^2} \\ 
              &=& \frac{\alpha}{2} \int \frac{(1+u +1-u) \mathrm{d}u}{(1-u)(1+u)} \\
              &=& \frac{\alpha}{2} \int \left ( \frac{1}{1-u} + \frac{1}{1+u} \right ) \mathrm{d}u \\
              &=& \frac{\alpha}{2} ( \log(1+u) - \log(1-u) ).
 \end{eqnarray}

By following a similar approach, we then obtain solutions for $I_{-3}$ and $I_{-5}$.

\subsection*{Integration in the 2D case}

The kernel function that we have chosen for the case of 2D space is analogous to that for 3D space and is given by:

\begin{equation}
   W(r) = \frac{10}{7h^2\pi}  \begin{cases}
      1 - 1.5 \left( \frac{r}{h} \right )^2 + 0.75 \left ( \frac{r}{h} \right )^3, & r\leq h; \\
      0.25 \left (2 - \left (\frac{r}{h} \right ) \right )^3, & h\leq r\leq 2h; \\
      0, & r\geq 2h. \\
  \end{cases} 
\end{equation}

By applying Green's Theorem, as stated in equation \ref{greenthm}, we can construct function $\mathbf{H} = H_{r} \mathbf{\hat{r}}$, such that $\frac{1}{r} \frac{\partial (r H_{r})}{\partial r} = W(r)$. As we integrate, we obtain the following solution for $H_r$:

\begin{equation}
   H_r(r) = \frac{1}{r} \frac{5}{7h^2\pi}  \begin{cases}
      r^2 - \frac{3}{4 h^2} r^4 + \frac{3}{10 h^3} r^5, & r\leq h; \\
      2 r^2 - \frac{2}{h} r^3 + \frac{3}{4 h^2} r^4 - \frac{1}{10 h^3} r^5 - \frac{1}{10} h^2, & h\leq r\leq 2h; \\
      \frac{7}{10} h^2, & r\geq 2h, \\
  \end{cases} \label{eq:torus}
\end{equation}

Finally, by expressing $r=\frac{r_0}{\mu}$ and integrating further, we have the solution:

\begin{equation}
   I_{P}=\int H_r(r) r \mathrm{d}\phi = \frac{5 r_0^2}{7h^2\pi}  \begin{cases}
      I_{-2} - \frac{3}{4} \left (\frac{r_0}{h} \right )^2 I_{-4} + \frac{3}{10} \left (\frac{r_0}{h} \right )^3 I_{-5}, & r\leq h; \\
       & \\
      2 I_{-2} - 2 \frac{r_0}{h} I_{-3} + \frac{3}{4} \left (\frac{r_0}{h} \right )^2 I_{-4} - \frac{1}{10} \left (\frac{r_0}{h} \right )^3 I_{-5} - \frac{1}{10} \left (\frac{r_0}{h} \right )^{-2 }I_0, & h\leq r\leq 2h; \\
       & \\
      \frac{7}{10} \left (\frac{r_0}{h} \right )^{-2} I_0, & r\geq 2h, \\
  \end{cases} \label{eq:torus}
\end{equation}

where $I_0$, $I_{-2}$, $I_{-3}$, $I_{-4}$, $I_{-5}$ are defined as previously.

\end {document}